\documentclass[aps,prd,twocolumn,superscriptaddress]{revtex4-2}

\usepackage[dvips]{graphicx}
\usepackage{ulem}
\usepackage[usenames]{color}
\usepackage{amsmath}
\usepackage{amssymb}
\usepackage{bm}
\usepackage{mathrsfs}
\usepackage{amsbsy}
\usepackage{hyperref}
\usepackage[all]{hypcap}
\usepackage{longtable}
\usepackage{url}
\usepackage{multirow}

\begin{document}

\title{Mass spectra of doubly charmed tetraquarks $T_{cc}$}

\author{You-You Lin}

\author{Ji-Ying Wang}

\author{Ailin Zhang}
\email{zhangal@shu.edu.cn}
\affiliation{Department of Physics, Shanghai University, Shanghai 200444, China}

\begin{abstract}
Motivated by the first observation of a doubly charmed tetraquark candidate $T_{cc}(3875)^+$, we perform a systematic calculation of the masses of some constituent diquarks and the doubly charmed tetraquark states from $1S$ to $2P$ excitations in a nonrelativistic constituent quark potential model. First, the mass of the charmed spin-$1$ diquark is predicted with $\sim 3500$ MeV, and the masses of the light ``good''(spin-$0$ and isospin-$0$) and ``bad''(spin-$1$ and isospin-$1$) antidiquark are predicted with $\sim 670$ MeV and $\sim 840$ MeV, respectively. The mass difference between a light ``good'' and a light ``bad'' diquark is $\sim 170$ MeV, which is consistent with previous phenomenological analyses. The interaction between the charmed diquark and the light antidiquark is then modulated by the quark-antiquark potential in the model, and the mass spectra of doubly charmed tetraquarks from $1S$ to $2P$ excitations are calculated with the obtained masses of diquarks/antidiquarks and refitted parameters from $T_{cc}(3875)^+$ and $X(3872)$. The pattern of the mass spectra of doubly charmed tetraquarks is obtained explicitly. The $1S-1P$ and $1S-2S$ mass splittings of doubly charmed tetraquarks are about $410-430$ MeV and $610-650$ MeV, respectively, and this mass splittings pattern is similar to that for ordinary $D$ mesons. The mass splittings of doubly charmed tetraquarks with a ``good'' diquark is about $10-25$ MeV higher than the corresponding mass splittings of doubly charmed tetraquarks with a ``bad'' diquark from $1S$ to $2P$ excitations. When $T_{cc}(3875)^+$ is assumed as a $IJ^P=01^+$ ground doubly charmed tetraquark candidate with a light ``good'' antidiquark, the next $P$-wave excited doubly charmed tetraquarks are predicted to have masses with an average $420-430$ MeV higher. A $IJ^P=10^+$ ground doubly charmed tetraquark $T_{cc}(3834)$ located around $3834$ MeV is also expected.
\end{abstract}

\maketitle
\section{Introduction}\label{intro}
Over past two decades, numerous exotic hadrons have been found in experiments and the ``particle zoo'' is rapidly expanding. In particular, four quark states have raised physicist's great concern since the discovery of $X(3872)$. Allowed by QCD theory, there exist several kinds of four quark states: tetraquark state~\cite{Jaffe:1976ig,Jaffe:2003sg,Maiani:2004uc,Maiani:2004vq,Ebert:2007rn,Yang:2020atz} in diquark-antidiquark configuration, molecular state~\cite{Voloshin:1976ap,PhysRevD.27.588,Tornqvist:1991ks,Tornqvist:1993ng,Wong:2003xk,Barnes:2003dj,Hosaka:2016pey,Liu:2019stu,Feijoo:2021ppq} in meson-meson configuration, hadro-charmonium~\cite{Dubynskiy:2008mq,Wang:2013kra}, and adjoint charmonium~\cite{Braaten:2013boa} etc. In Ref.~\cite{Braaten:2014qka}, seven kinds of explanations of XYZ mesons have been listed though there is no compelling explanation for the pattern of XYZ mesons. A doubly charmed tetraquark has quark component: $cc\bar u\bar d$. As a possible partner of $X(3872)$, the first doubly charmed tetraquark candidate $T_{cc}(3875)^+$~\cite{LHCb:2021vvq,LHCb:2021auc} was reported by LHCb in a $D^0 D^0\pi^+$ invariant mass spectrum in proton-proton collisions in $2021$. $T_{cc}(3875)^+$ is located just below the $D^{*+} D^0$ mass threshold with a very narrow width. With a Breit-Wigner parametrization, its mass relative to the charmed meson pair is measured
\begin{equation}
\delta m_{BW}\equiv m_{BW}-(m_{D^{*+}}+m_{D^0})=-273\pm61\pm5^{+11}_{-14}~\mathrm{keV},
\end{equation}
where the charmed mesons have masses~\cite{ParticleDataGroup:2020ssz}
\begin{equation}
\begin{aligned}
 m_{D^{*+}}&=2010.26\pm0.05~\mathrm{MeV},\\
 m_{D^0}&=1864.84\pm0.05~\mathrm{MeV}.
\end{aligned}
\end{equation}

For the proximity of $T_{cc}(3875)^+$ to the $D^{*+} D^0$ mass threshold, a unitarised scheme incorporating the threshold effect is employed and the mass parameter is shifted to~\cite{LHCb:2021auc}
\begin{equation}
\delta m_{pole}=-360\pm40^{+4}_{-0}~\mathrm{keV}.
\end{equation}
Moreover, the spin-parity quantum numbers of the isoscalar $T_{cc}(3875)^+$ are in favor of $J^P=1^+$ in experiment.

So far, experimental data of doubly heavy tetraquark is scarce, but there are many theoretical studies of doubly charmed four quark states. In the 1980's, most studies of doubly charmed four quark states concerned on their stability and existence~\cite{Lipkin:1986dw,Carlson:1987hh,Semay:1994ht}. In recent years, as more experimental data of heavy baryons and multi-quarks accumulated, the mass spectra, decay widths and production rates of doubly charmed four quark states have been investigated with various methods~\cite{Vijande:2003ki,Feijoo:2021ppq,Albaladejo:2021vln,Wang:2022jop}, such as constituent quark models~\cite{Gelman:2002wf,Vijande:2003ki,Vijande:2007rf,Ebert:2007rn,Lee:2009rt,Kleiv:2013dta,Braaten:2014qka,Luo:2017eub,Karliner:2017qjm,Eichten:2017ffp,Richard:2018yrm,Braaten:2020nwp,Lu:2020rog,Cheng:2020wxa,Meng:2020knc,Qin:2020zlg,Weng:2021hje,Deng:2021gnb,Kim:2022mpa}, flux-tube model~\cite{Deng:2018kly}, QCD sum rule~\cite{Navarra:2007yw,Du:2012wp,Wang:2017dtg,Esau:2019hqw,Agaev:2022ast}, Bethe-Salpeter equation~\cite{Feng:2013kea,Ding:2020dio}, lattice QCD simulation~\cite{Ikeda:2013vwa,Junnarkar:2018twb,Padmanath:2022cvl}, etc. More references could be found in relevant reviews~\cite{Hosaka:2016pey,Lebed:2016hpi,Esposito:2016noz,Guo:2017jvc,Liu:2019zoy,Chen:2022asf} and references therein.

There are manifest bifurcations on the nature of the doubly charmed $T_{cc}(3875)^+$. There are generally two kinds of interpretations. As $T_{cc}(3875)^{+}$ has mass very close to $D^{*+} D^0$ mass threshold, the interpretation as a doubly charmed hadronic molecular state has been discussed extensively~\cite{Feijoo:2021ppq,Albaladejo:2021vln,Deng:2021gnb,Agaev:2022ast,Wang:2022jop}. However, the possibility as a doubly charmed tetraquark cannot be ruled out. The properties of $T_{cc}(3875)^+$ as a doubly charmed tetraquark in diquark-antidiquark or compact tetraquark picture have also been investigated~\cite{Weng:2021hje,Kim:2022mpa}.

Different interactions and models could be employed to compute the mass spectra of doubly charmed four quark states, and therefore the pattern of the mass spectra can exhibit much information about the interactions and structures inside hadrons. The calculation of mass spectra is an important way to detect the dynamics and structure in tetraquarks. In practice, the interpretation of four quark states candidates depends also heavily on the internal quark interactions and models.

There are four components in a doubly charmed tetraquark, and the internal interactions are complex. The mass spectra of the doubly charmed four quark states has been calculated in different approximations. In a four-body picture, the mass of ground doubly charmed tetraquark with isospin-$0$ is calculated with $3916$ MeV($1$ MeV below the lowest meson-meson threshold) based on a non-relativistic potential which reproduced very well the masses of mesons and baryons~\cite{Semay:1994ht}. A similar four-body computation with a modified potential indicates that the ground doubly charmed four quark state has mass $23$ MeV below the $D^{*+} D^0$ threshold~\cite{Meng:2020knc}. In a three-body picture, the mass of ground doubly charmed four quark state is obtained with $3961$ MeV in a chiral diquark model~\cite{Kim:2022mpa}, where two light antiquarks make a point-like antidiquark, and then the antidiquark interacts with other two charmed quarks. In a two-body picture, the mass spectra of the doubly charmed four quark states was also investigated in a relativistic potential model~\cite{Ebert:2007rn}, where the diquark mass is calculated with the same potential between the diquark and the antidiquark. The ground doubly charmed tetraquark with $IJ^P=01^+$ is predicted with $3935$ MeV. Systemic computation of the mass spectra of doubly charmed tetraquarks including high excitations is scarce and urgently required.

Though the internal structure of a tetraquark is not clear, many studies indicate that the diquark and antidiquark is a good stuff to construct the tetraquark. A diquark was first mentioned by Gell-Mann~\cite{GellMann1964}, and prevailed in multi-quark hadronic systems as a fundamental antisymmetric composite of two strongly correlated quarks~\cite{Ida:1966ev,Lichtenberg:1967zz,Anselmino:1992vg,Jaffe:2003sg,Karliner:2003dt,Maiani:2004uc,Maiani:2004vq,Jaffe:2004ph,Maiani:2015vwa,Chen:2016qju,Lebed:2016hpi,Maiani:2017kyi,Olsen:2017bmm,Barabanov:2020jvn}.
It is also pointed out that a charmed diquark is indispensable in the construction of the doubly charmed baryons and tetraquarks in many investigations~\cite{Navarra:2007yw,Feng:2013kea,Eichten:2017ffp,Karliner:2017qjm,Richard:2018yrm,Braaten:2020nwp,Cheng:2020wxa}.


In nature, it is possible that a doubly charmed tetraquark consists of a fundamental charmed diquark and light antidiquark where two charmed quarks make a compact heavy diquark and two light quarks make a compact light antidiquark. On the other hand, the diquark-antidiquark picture of the tetraquark is a good approximation for simplicity. It is expected that the calculation of the mass spectra of doubly charmed tetraquarks $[cc][\bar u \bar d]$ in the diquark-antidiquark picture may discover more information on the dynamics and structure in tetraquark. Another aim is to show the pattern of mass spectra of doubly charmed tetraquarks around $4$ GeV, and to provide more information for the search of tetraquark in experiments.

For these purposes, the potential proposed by Semay et al~\cite{Semay:1994ht,Silvestre-Brac:1996myf} is employed in our calculation, similar potential has already been employed in some studies of doubly charmed tetraquarks~\cite{Semay:1994ht,Richard:2018yrm,Meng:2020knc,Meng:2021yjr,Kim:2022mpa}. In contrast with normal treatment of tetraquarks in terms of diquarks and antidiquarks, masses of the charmed diquark and light antidiquarks are obtained through a quark-quark interaction instead of parameters input. Subsequently, the mass spectra of doubly charmed tetraquark is calculated through a similar quark-antiquark interaction.

The work is organized as follows, the construction of the potential model and the wave functions is presented in Sec.~\ref{model} after the introduction. The numerical calculation and the mass spectra are given in Sec.~\ref{results}. The last section is devoted to a brief summary.

\section{Constituent diquark-antidiquark potential model}\label{model}
\subsection{Potential Model}\label{ss:pm}
To describe the mass spectra of mesons more accurately, based on the familiar ``Coulomb+linear'' potentials, Semay et al proposed a modified potential for quark-antiquark interaction in the following explicit form in a norelativistic potential model~\cite{Semay:1994ht,Silvestre-Brac:1996myf}
\begin{equation}\label{eq:brac}
\begin{aligned}
V_{q\bar q}=&V_0+(\vec{s}_1 \cdot \vec{s}_2)V_{ss}\\
=&-\frac{\alpha(1-e^{-r/r_c})}{r}+\lambda r^{p}+C\\
&+(\vec{s}_1 \cdot \vec{s}_2)\frac{8\kappa(1-e^{-r/r_c})}{3m_1 m_2\sqrt{\pi}}\frac{e^{-r^2/r_0^2}}{r_0^3},
\end{aligned}
\end{equation}
where $r_0$ is determined by
\begin{equation}\label{eq:r0}
	r_0=A{\left(\frac{2m_1m_2}{m_1+m_2}\right)}^{-B},
\end{equation}
and the detail of other parameters can be found in Refs.~\cite{Semay:1994ht,Silvestre-Brac:1996myf}. In baryon systems with heavy quark, the mass spectra of baryons are also well described by the potential. Of course, the quark-quark interaction potential in baryon or multiquark states is obtained by a $V_{qq}={\frac{1}{2}}V_{q\bar q}$ prescription.

The parameters $r_c$ and $p$ indicate the deviation of the potential from the Cornell potential at short and long distance, respectively, and they may take different values in mesons and baryons~\cite{Semay:1994ht,Silvestre-Brac:1996myf}. Four kinds of potentials were proposed based on the two parameters:
\begin{equation}
\begin{aligned}
AL1~~ \mathrm{potential}:&~p=1,~r_c=0;\\
AL2~~ \mathrm{potential}:&~p=1,~r_c \neq 0;\\
AP1~~ \mathrm{potential}:&~p=\frac{2}{3},~r_c=0;\\
AP2~~ \mathrm{potential}:&~p=\frac{2}{3},~r_c \neq 0,
\end{aligned}
\end{equation}
where the letter $L$ or $P$ denotes the linear ($p=1$) or the ${\frac{2}{3}}$-Power ($p={\frac{2}{3}}$) confinement, respectively. The parameter $p={\frac{2}{3}}$ gives the correct asymptotic (for large angular momentum) Regge trajectories of mesons and baryons~\cite{FabreDeLaRipelle:1988zr,Semay:1994ht}. The number $1$ or $2$ indicates two types of the parameter $r_c$ (zero or finite number) where $r_c=0$ means that $\alpha$ and $\kappa$ are constant parameters.

The spectra of baryons and mesons from these kinds of potentials are better described than the common Coulomb-plus-linear potential~\cite{Semay:1994ht,Silvestre-Brac:1996myf}. These potentials are employed in our calculation of the masses of diquarks/antiquarks and tetraquarks. Since the diquark or antidiquark is assumed as a compact component where there does not exist an angular momentum between the quarks or antiquarks, the $AL$ potentials are employed. And only the tetraquarks from $S-$wave excitations to $P-$wave excitations are concerned in our calculation, the $AL$ potentials are also preferred.

Thus the full Hamiltonian of the doubly charmed tetraquark is constructed as
\begin{equation}\label{eq:ham}
H=m_1+m_2+\frac{p^2}{2\mu}+V_{al}+V_{sl},
\end{equation}
where $m_{1}$ and $m_{2}$ are the mass of the charmed diquark and the light antidiquark, respectively. $\frac{p^2}{2\mu}$ is the kinetic energy of the tetraquark in c.m.s, determined by the relative momentum $p$ and reduced mass $\mu=\frac{m_1 m_2}{m_1+m_2}$. $V_{al}$ is the $AL$-type potential~\cite{Silvestre-Brac:1996myf} in Eq. (\ref{eq:brac}).

The last term $V_{sl}$ in Eq.~(\ref{eq:ham}) represents the coupling between the spin and the orbital angular momentum. A universal Breit-Fermi approximation is employed as~\cite{DeRujula:1975qlm}
\begin{equation}\label{eq:vsl}
\begin{aligned}
V_{sl}=&V_{so}+V_{ten}=\frac{\alpha(1-e^{-r/r_c})}{r^3}\left(\frac{1}{m_1}+\frac{1}{m_2}\right)\\
&\times\left(\frac{\vec{s}_1}{m_1}+\frac{\vec{s}_2}{m_2}\right) \cdot \vec{L}-\frac{1}{2r}\frac{\partial V_{conf}}{\partial r}\left(\frac{\vec{s}_1}{m_1^2}+\frac{\vec{s}_2}{m_2^2}\right) \cdot \vec{L}\\
&+\frac{1}{3m_1 m_2}\left(\frac{1}{r}\frac{\partial V_{Coul}}{\partial r}-\frac{\partial^2V_{Coul}}{\partial r^2}\right)\\
&\times\left(\frac{3\vec{s}_1 \cdot \vec{r} \vec{s}_2  \cdot \vec{r}}{r^2}-\vec{s}_1 \cdot \vec{s}_2\right),	
\end{aligned}
\end{equation}
where $V_{Coul}$ is the modified Coulomb potential in this work, thus the explicit form follows
\begin{equation}\label{eq:vsle}
\begin{aligned}
V_{sl}=&V_{1}{\vec{L} \cdot \vec{s}_1}+V_{2}{\vec{L} \cdot \vec{s}_2}+V_t\left(\frac{3\vec{s}_1 \cdot \vec{r} \vec{s}_2  \cdot \vec{r}}{r^2}-\vec{s}_1 \cdot \vec{s}_2\right)	\\
=&\left[\frac{\alpha(1-e^{-r/r_c})}{m_1 m_2 r^3}+\frac{1}{m_1^2}\left(\frac{\alpha(1-e^{-r/r_c})}{2r^3}-\frac{\lambda}{2r}\right)\right]{\vec{L} \cdot \vec{s}_1}\\
&+\left[\frac{\alpha(1-e^{-r/r_c})}{m_1 m_2 r^3}+\frac{1}{m_2^2}\left(\frac{\alpha(1-e^{-r/r_c})}{2r^3}-\frac{\lambda}{2r}\right)\right]\\
&\times{\vec{L} \cdot \vec{s}_2}+\frac{\alpha(1-e^{-r/r_c})}{m_1m_2 r^3}\left(\frac{3\vec{s}_1 \cdot \vec{r} \vec{s}_2  \cdot \vec{r}}{r^2}-\vec{s}_1 \cdot \vec{s}_2\right),
\end{aligned}
\end{equation}
where $\vec{s}_1$ and $\vec{s}_2$ denote the spin angular momentum of the charmed diquark and the light antidiquark, respectively, and $\vec{L}$ stands for their orbital angular momentum. $r_c=0$ and $r_c\neq 0$ corresponding to the $AL1$ and $AL2$ potential, respectively, will be taken used of.

In order to compute the masses of the charmed diquark and light antidiquarks, similar equations as Eq.~(\ref{eq:ham}) hold except that the quark-quark potential inside the charmed diquark and light antidiquark has the forms as following
\begin{equation}\label{eq:vcc}
\begin{aligned}
V_{qq}=&\frac{1}{2}\left[-\frac{\alpha(1-e^{-r/r_c})}{r}+\lambda r+C\right.\\
&\left.+(\vec{s}_q \cdot \vec{s}_q)\frac{8\kappa(1-e^{-r/r_c})}{3m_q m_q\sqrt{\pi}}\frac{e^{-r^2/r_0^2}}{r_0^3}\right].
\end{aligned}
\end{equation}
As the potential is proportional to the product of two Gell-Mann matrices, $\vec{\lambda}_i \cdot \vec{\lambda}_j$, the potential between two quarks is half the potential between a quark and an antiquark when the diquark is of color antitriplet.

\subsection{Wave Function}
In the diquark-antidiquark picture, the total wave function of a doubly charmed tetraquark is composed of color, flavor, spin and spatial parts,
\begin{equation}\label{eq:wf}
	\Psi_{JM}=\chi_f\otimes\chi_c\otimes[\Psi_{lm}(\vec{r})\otimes \chi_s]_{JM}.
\end{equation}
$\chi_f$ is the flavor wave function
\begin{equation}\label{eq:flavor}
\chi_{f}=|\{cc\}(\bar{u}\bar{d})\rangle,
\end{equation}
where the curly bracket outside the charmed diquark represents its symmetric flavor wave function, and there are two possibilities for the light antidiquark with different spin and isospin. In the following, a square bracket outside the $\bar u\bar d$ antidiquark indicates the antisymmetric flavor structure with a spin and isospin $0$, while a curly bracket outside the $\bar u\bar d$ antidiquark indicates the symmetric flavor structure with a spin and isospin $1$. They are the so called ``good'' and ``bad'' antidiquarks in Ref.~\cite{Jaffe:2004ph}, respectively, which are sometimes denoted with the ``scalar'' and ``vector'' antidiquarks directly. In what follows, these two notations will be employed for light antidiquarks everywhere. As a diquark is assumed without spatial structure, the wave function of a diquark consists of the color, spin and flavor wave function. Thus four kinds of color-spin wave functions are allowed in the $\mathbf{\bar{3}}_c\otimes\mathbf{3}_c$ color configuration
\begin{flalign}
    & \chi_{3,1}=|\{cc\}^{\bar 3}_1[\bar{u}\bar{d}]^3_0\rangle_1,\label{eq:cs1}\\
	&\chi_{3,0}=|\{cc\}^{\bar 3}_1\{\bar{u}\bar{d}\}^3_1\rangle_0,\label{eq:cs2}\\
	&\chi_{3,1}=|\{cc\}^{\bar 3}_1\{\bar{u}\bar{d}\}^3_1\rangle_1,\label{eq:cs3}\\
	&\chi_{3,2}=|\{cc\}^{\bar 3}_1\{\bar{u}\bar{d}\}^3_1\rangle_2,\label{eq:cs4}
\end{flalign}
where the notations $|\{cc\}^{color}_{spin}(\bar{u}\bar{d})^{color}_{spin}\rangle_{total~spin}$ were employed. The first wave function corresponds to the tetraquark composed of ``good'' diquark, while the other three are those with ``bad'' diquark.

Similarly, two kinds of color-spin wave functions are allowed in the $\mathbf{6}_c\otimes \mathbf{\bar{6}}_c$ color configuration for the tetraquark,
\begin{flalign}
    & \chi_{6,1}=|\{cc\}^6_0[\bar{u}\bar{d}]^{\bar{6}}_1\rangle_1,\label{eq:cs61}\\
	&\chi_{6,0}=|\{cc\}^6_0\{\bar{u}\bar{d}\}^{\bar{6}}_0\rangle_0.\label{eq:cs62}
\end{flalign}
However, the color interactions within these kinds of configurations are repulsive~\cite{Jaffe:2004ph}, and thus these diquarks/antidiquarks are not likely to make compact correlated clusters. Therefore, the configurations in the $\mathbf{6}_c$ color representation are not discussed in this work.


The spatial wave function $\psi_{lm}(\vec{r})$ is obtained through a Gaussian expansion method(GEM)~\cite{Hiyama:2003cu}, in which the wave function is expanded in terms of Gaussian functions as the trial wave function in Schr\"{o}dinger equation,
\begin{align}
\Psi_{lm}(\vec{r})&=\sum_{n=1}^{n_{max}}c_{nl}\phi_{nlm}(\vec{r}),\label{eq:plm}\\
\phi_{nlm}(\vec{r})&=N_{nl}r^l e^{-r^2/r_n^2}Y_{lm}(\hat{r}),\label{eq:pnlm}
\end{align}
$N_{nl}$ is the normalization coefficient of the state with radial and orbital angular momentum quantum numbers $n$ and $l$. The size parameter $r_n$ is obtained in a geometric progression
\begin{equation}\label{eq:rn}
	r_n=r_1a^{n-1}(n=1,...,n_{max}),
\end{equation}
from three parameters $\{n_{max},~r_1,~r_{n_{max}}\}$~\cite{Hiyama:2003cu}.

Through the Rayleigh-Ritz variational principle, the Schr\"odinger equation is transformed to the following matrix eigenvalue problem
\begin{equation}\label{eq:gem}
\sum_{n=1}^{n_{max}} \sum_{n^{\prime}=1}^{n_{max}}(H_{nn^{\prime}}-E_{nl}N_{nn^{\prime}})c_{nl}=0,
\end{equation}
where $H_{nn^{\prime}}$, $N_{nn^{\prime}}$ are the matrix element and the overlap integral,
\begin{align}
H_{nn^{\prime}}&=\langle\phi_{n^{\prime}lm}(\vec{r})|H|\phi_{nlm}(\vec{r})\rangle,\label{eq:hnn}\\
N_{nn^{\prime}}&=\langle\phi_{n^{\prime}lm}(\vec{r})|\phi_{nlm}(\vec{r}\rangle.\label{eq:nnn}
\end{align}
The expansion coefficients $c_{nl}$ in Eq.~(\ref{eq:plm}) and the eigenenergy $E$ are then obtained. To obtain convergent results, we fix $r_0$ with $0.1~\mathrm{fm}$ as a start, which affects little on the final results, and modify the values of $r_n$ and $n_{max}$ as did in Refs.~\cite{Deng:2021gnb,Chen:2022ros}. In our geometric progression, the variational parameters $\{r_1,r_{n_{max}},n_{max}\}$ are finally fixed as $\{0.1~\mathrm{fm},5~\mathrm{fm},25\}$.

\subsection{Matrix Elements}
After the diquark masses have been computed, the mass spectra of the tetraquark states is obtained from Eq.~(\ref{eq:ham}). To obtain the contributions of spin-dependent terms, we first expand the wave function in the $LS$ coupling scheme. That is to say, the spin angular momentum of the diquark and the antidiquark couples to $\vec{s}_1+\vec{s}_2=\vec{S}$, then couple with the orbital angular momentum $\vec{L}$ to get the total angular momentum $\vec{S}+\vec{L}=\vec{J}$ of tetraquark state which is denoted as $|^{2S+1}L_J\rangle$. The mass matrix of the spin-dependent interaction is expressed as
\begin{equation}
H_{SD}=\begin{pmatrix}H_{11}&H_{12}& \cdots\\
H_{21}&H_{22}& \cdots\\
\vdots & \vdots & \ddots\end{pmatrix}.
\end{equation}
Similar to the heavy-light meson, there are mixings between $[cc][\bar u\bar d]$ states with different total spin but the same total angular momentum, $|^{2S+1}L_J\rangle \leftrightarrow  |^{2S^\prime+1}L_J\rangle$ resulted from the spin-orbit interaction~\cite{Liu:2015lka}.

For the $S-$wave and $P-$wave excited doubly charmed tetraquarks without radial excitation, according to the wave functions, there are one $S-$wave and three $P-$wave excited tetraquarks made from a charmed diquark with spin-$1$ and a light ``good'' antidiquark. There are three $S-$wave and seven $P-$wave excited tetraquarks made from a charmed diquark with spin-$1$ and a light ``bad'' antidiquark.

For the seven $P-$wave excited tetraquarks with light ``bad'' antidiquark, the mass matrix elements are expressed in $|^{2S+1}L_J\rangle$ basis as follows:

(i) For the states $^3P_0$ and $^5P_3$ with $S=1$ and $S=2$, no mixing is expected, and the contributions of the spin-dependent terms are
\begin{flalign}
&J=0:H_{SD}=-V_{ss}-V_{1}-V_{2}-2V_t,\label{eq:hs0}\\
&J=3:H_{SD}=V_{ss}+V_{1}+V_{2}-\frac{2}{5}V_t.\label{eq:hs3}	
\end{flalign}

(ii) For $J=1$ states, there are mixing among $^1P_1$, $^3P_1$ and $^5P_1$. The mass matrix elements are
\begin{flalign}
&H_{11}=-2V_{ss},\label{eq:1h11}\\
&H_{12}=H_{21}^{*}=\frac{2}{\sqrt{3}}V_{1}-\frac{2}{\sqrt{3}}V_{2},\label{eq:1h12}\\
&H_{13}=H_{31}^{*}=\frac{2}{\sqrt{5}}V_t,\label{eq:1h13}\\
&H_{22}=-V_{ss}-\frac{1}{2}V_{1}-\frac{1}{2}V_{2}+V_t,\label{eq:1h22}\\
&H_{23}=H_{32}^{*}={\frac{\sqrt{15}}{6}}V_{1}-{\frac{\sqrt{15}}{6}}V_{2},\label{eq:1h23}\\
&H_{33}=V_{ss}-\frac{3}{2}V_{1}-\frac{3}{2}V_{2}-\frac{7}{5}V_t.\label{eq:1h33}
\end{flalign}

(iii) For $J=2$ states, there are mixing between $^3P_2$ and $^5P_2$. The mass matrix elements are
\begin{flalign}
&H_{11}=-V_{ss}+\frac{1}{2}V_{1}+\frac{1}{2}V_{2}-\frac{1}{5}V_t,\label{eq:2h11}\\
&H_{12}=H_{21}^{*}=\frac{\sqrt{3}}{2}V_{1}-\frac{\sqrt{3}}{2}V_{2},\label{eq:2h12}\\
&H_{22}=V_{ss}-\frac{1}{2}V_{1}-\frac{1}{2}V_{2}+\frac{7}{5}V_t.
\label{eq:2h22}	
\end{flalign}

The $V_{ss}$, $V_{1}$, $V_2$, $V_t$ are the spin-spin, spin-orbit and tensor interaction potentials displayed in Sec.~\ref{ss:pm}.

\section{Numerical results}\label{results}
\subsection{Model Parameters}
The mass of constituent quarks and the parameters in the $AL$ quark-quark potentials for the calculation of charmed diquark and light antidiquarks are taken as those in Refs.~\cite{Semay:1994ht,Silvestre-Brac:1996myf} where the mass spectra of mesons and baryons is simultaneously reproduced well. In Table~\ref{tab:para}, the parameters adopted in our computation are presented.
\begin{table}[htb]
\caption{\label{tab:para}
Parameters in two $AL$ potentials.}
\begin{ruledtabular}
\begin{tabular}{ccccccccc}
Parameters &$AL1$ & $AL2$ \\
 \colrule
$m_{u,d}(\mathrm{GeV})$  & $0.315$   & $0.320$   \\
$m_c(\mathrm{GeV})$     & $1.836$   & $1.851$   \\
$\alpha$     & $0.5069$  & $0.5871$  \\
$\kappa$    & $1.8609$  & $1.8475$  \\
$\lambda(\mathrm{GeV}^2)$ & $0.1653$  & $0.1673$  \\
$C(\mathrm{GeV})$   & $-0.8321$ & $-0.8182$ \\
$B$     & $0.2204$  & $0.2132$  \\
$A(\mathrm{GeV}^{B-1})$    & $1.6553$  & $1.6560$  \\
$r_c(\mathrm{GeV^{-1}})$    & $0$       & $0.1844$
\end{tabular}
\end{ruledtabular}
\end{table}

With these parameters, the masses of relevant diquarks and antidiquarks are computed with the $AL$ potentials, and the results are presented in Table~\ref{tab:mdd}. The light ``good'' antidiquark has a mass $\sim 170$ MeV lower than the light ``bad'' antidiquark. This result is consistent with the phenomenological analysis in Ref.~\cite{Jaffe:2004ph}, where the diquark mass differences between ``good'' and ``bad'' diquarks is $\sim 200$ MeV. In particular, our obtained diquark mass difference is spectator quark independent while the diquark mass difference extracted from phenomenological analysis is spectator quark dependent.

The mass of the vector charmed diquark is almost the same as the predicted $3.51$ GeV in QCD sum rule~\cite{Esau:2019hqw}, while the masses of the light antidiquarks are larger than those fitted by phenomenological analysis~\cite{Maiani:2004vq} or calculated with QCD sum rule~\cite{Zhang:2006xp}. However, the constituent mass difference of the light ``good'' antidiquark and the light antiquark, $m_{\bar u\bar d}-m_{\bar u,\bar d}\sim 350$ MeV, is only $40$ MeV larger than the spectator quark dependent $\sim 310$ MeV extracted from phenomenological analysis~\cite{Jaffe:2004ph,Maiani:2004vq}.
\begin{table}[htb]
\caption{\label{tab:mdd}Mass (in MeV) of the charmed diquark and light antidiquarks in $AL$ potentials.}
\begin{ruledtabular}
\begin{tabular}{cccccc}
  & spin & $AL1$  & $AL2$ \\
\colrule
$m_{cc}$     &  $s=1$    & $3499.5$ & $3518.6$ \\
\multirow{2}{*}{$m_{\bar u\bar d}$} & $s=0$ & $666.2$  & $673.9$  \\
                          & $s=1$ & $834.1$  & $841.8$
\end{tabular}
\end{ruledtabular}
\end{table}

Smaller masses of the ground doubly and hidden charmed tetraquarks are obtained when we proceed the calculation in terms of the parameters in Table~\ref{tab:para}. The obtained masses of the ground doubly and hidden charmed tetraquarks are about $130$-$160$ MeV smaller than the masses of $T_{cc}(3875)$ and $X(3872)$.
The predicted smaller masses may result from the inner structure of tetraquarks. As known, for the cluster structure of diquarks/antidiquarks and a relative larger separation between a diquark and an antidiquark in comparison to the distance between a quark and an antiquark in ordinary mesons, the interaction intensity and accordingly the parameters in the $AL$ quark-antiquark potential between the charmed diquark and the light antidiquark may not the same as those in ordinary mesons. In Ref.~\cite{Weng:2021hje}, two schemes of parameters are adopted to study the $(qq)(\bar Q\bar Q)$ tetraquarks for the difference of spatial configurations in the conventional hadrons and tetraquarks. In this work, the parameters $\alpha$ and $\lambda$ are refixed through the two observed tetraquark candidates $T_{cc}(3875)^+$ and $X(3872)$.

There are many explanations of $X(3872)$ and $T_{cc}(3875)$, but we are far from understanding their structure and dynamics. In Ref.~\cite{Weng:2021hje}, a lowest $IJ^P=01^+$ $T_{cc}$ with $(qq)(\bar c\bar c)$ configurations is predicted with mass $3869$ MeV in their scheme II, and in Ref.~\cite{Kim:2022mpa}, a lowest $J^P=1^+$ $T_{cc}$ with $(cc)(\bar q\bar q)$ configurations with the flavor $3$ antidiquarks is predicted with mass $\sim 3961$ MeV. In contrast to mostly advocated molecule interpretation, we think it is rational to assume $T_{cc}(3875)^+$ as a ground $IJ^P=01^+$ doubly charmed tetraquark.

In recently, a measurement of the radiative decay ratio $\frac{\Gamma_{\chi_{c1}(3872)\to \Psi(2S)\gamma}}{\Gamma_{\chi_{c1}(3872)\to J/\Psi\gamma}}$ does not support the popular molecule interpretation of $X(3872)$~\cite{Grinstein:2024rcu,LHCb:2024tpv}. In particular, $X(3872)$ is assumed to have $c\bar c$ charmonium component in some analyses. Obviously, the mixing will complex the behaviors of $X(3872)$ and would make the mass of the pure hidden charmed tetraquark deviate from the mass of $X(3872)$.
In this paper, we will not focus on the mixing before some detailed mixing information has been discovered by experiments. Instead, the mixing effect will be taken into account through some uncertainties of the masses of $X(3872)$ and $T_{cc}(3875)$. $X(3872)$ is assumed as a ground $J^{PC}=1^{++}$ hidden charmed tetraquark made from a $cq$ diquark and a $\bar c\bar q$ antidiquark in our study~\cite{Lin}.

In fact, the explanation of exotic states depends heavily on the quark dynamics and employed model. We hope this prescription could make a good description of the doubly charmed tetraquarks. The spin-spin interaction does not contribute to the mass of these two ground states, and the values of $\kappa$, $C$, $B$, $A$ and $r_c$ follow the original sets in Refs.~\cite{Semay:1994ht,Silvestre-Brac:1996myf} with uniformity.

There are two types of $AL$ potentials in the calculation of the masses of diquark/antidiquark and tetraquark. Once the charmed diquark and light antidiquark are computed with the same $r_c=0$ or $r_c\neq 0$ simultaneously, four sets of parameters for $\alpha$ and $\lambda$ are fixed. Hereafter for simplicity, they are denoted as Set I to Set IV. The obtained parameters of Set I to Set IV resulted from potentials $AL1$ and $AL1$-type, $AL2$ and $AL1$-type, $AL1$ and $AL2$-type, $AL2$ and $AL2$-type, respectively, in the diquark and tetraquark are presented in Table~\ref{tab:pal}. The fitted $\alpha$ (modified Coulomb term coefficient) in tetraquarks is smaller than that in ordinary mesons, while the fitted $\lambda$ (modified confinement term coefficient) is larger than that in ordinary mesons.
\begin{table}[htb]
\caption{\label{tab:pal}
Parameters of Set I to Set IV.}
\begin{ruledtabular}
\begin{tabular}{ccccccccc}
Set & I & II & III & IV\\
\colrule
$\alpha$ & $0.3910$ & $0.4205$ & $0.4290$ & $0.4600$\\
$\lambda(\mathrm{GeV}^2)$ & $0.1941$  & $0.1916$ & $0.1957$ & $0.1932$
\end{tabular}
\end{ruledtabular}
\end{table}

The isospin breaking and mixing effects play an important role in molecule states, where bosons are exchanged between color-singlet hadrons to bind hadronic molecules both in constituent quark potential model and in chiral Lagrangian theory~\cite{Sun:2024wxz,Li:2012cs,Lebed:2024zrp,Takeuchi:2014rsa,Dai:2023cyo}. The long-distance isospin dependence of the interactions in molecular models follows primarily from the isospin content of exchanged bosons. The isospin breaking and mixing effects may play an essential role in diquark-antidiquark states. In Refs.~\cite{Brodsky:2014xia,Lebed:2015tna,Sun:2024wxz}, an isospin-dependent interaction was introduced in the diquark-antidiquark picture for the hidden-charm tetraquarks, which brings in less than $10$ MeV to the mass spectra. This isospin-dependent interaction provides a dynamical explanation of the isospin breaking or mixing effects. In Ref.~\cite{Lebed:2024zrp}, a mixing scheme of the molecule and diquark-antidiquark is proposed, and the effects of $D^*D$ isospin is well analyzed.

To take into account the mixing and isospin breaking effects on the spectrum of tetraquarks, we vary the mass of $X(3872)$ with $50$ MeV around its experimental data, and our fitting procedure indicates that the parameters $\alpha$ and $\lambda$ decrease small with the increase of the mass. The $\alpha$ varies about $0.2$, and the $\lambda$ varies about $0.06$. We also vary the mass of $T_{cc}(3875)^+$ with $10$ MeV around its experimental data in the fitting procedure, and the parameters vary little and increase with the increase of the mass. The $\alpha$ varies about $0.02$, the $\lambda$ varies about $0.01$.

\subsection{Mass Spectra of doubly charmed tetraquarks}
With those parameters and the computed masses of diquarks and antidiquarks, the mass spectra of doubly charmed tetraquarks from $1S$ to $2P$ excitations are computed. The results of the doubly charmed tetraquarks composed of light ``good'' and ``bad'' antidiquark are presented in Tables~\ref{tab:m10l} and \ref{tab:m11l}, respectively.

In these tables, the first column gives all the allowed tetraquark states from $1S$- to $2P$-wave excitations with the spectroscopic notations in $LS$ coupling scheme. The spin-parity quantum numbers $J^{P}$ are listed in the second column with the parity $P=(-1)^L$. The third to the sixth columns present the obtained mass spectra with the refitted parameters Sets I-IV, respectively. For a comparison, the mass spectra in the chiral diquark model~\cite{Kim:2022mpa} is presented in the last column of Tables~\ref{tab:m10l} and \ref{tab:m11l}.



\begin{table*}[htb]
\caption{\label{tab:m10l}Mass spectra (in MeV) of doubly charmed tetraquark in $\mathbf{\bar{3}}_c\otimes\mathbf{3}_c$ color configuration with light ``good'' antidiquark.}
\begin{ruledtabular}
\begin{tabular}{cccccccc}
$n^{2S+1}L_J$ & $J^{P}$ & Set I & Set II & Set III & Set IV  & Ref.~\cite{Kim:2022mpa}\\
\colrule
$1^3S_1$ & $1^+$ & $3874.83$ & $3874.78$ & $3875.42$ & $3874.84$ & $3961$ \\
$1^3P_0$ & $0^-$ & $4274.51$ & $4276.50$  & $4278.91$ & $4280.18$ & $4253$ \\
$1^3P_1$ & $1^-$ & $4284.51$ & $4287.53$ & $4290.24$ & $4292.61$ & $4253$ \\
$1^3P_2$ & $2^-$ & $4304.50$  & $4309.58$ & $4312.90$  & $4317.47$ & $4253$ \\
$2^3S_1$ & $1^+$ & $4511.38$ & $4511.17$ & $4522.19$ & $4521.77$ & $4363$ \\
$2^3P_0$ & $0^-$ & $4792.06$ & $4790.46$ & $4802.52$ & $4800.42$ &        \\
$2^3P_1$ & $1^-$ & $4802.07$ & $4801.38$ & $4813.60$  & $4812.45$ &        \\
$2^3P_2$ & $2^-$ & $4822.07$ & $4823.23$ & $4835.74$ & $4836.52$ &
\end{tabular}
\end{ruledtabular}
\end{table*}

\begin{table*}[htb]
\caption{\label{tab:m11l}
Mass spectra (in MeV) of doubly charmed tetraquark in $\mathbf{\bar{3}}_c\otimes\mathbf{3}_c$ color configuration with light ``bad'' antidiquark.}
\begin{ruledtabular}
\begin{tabular}{ccccccc}
$n^{2S+1}L_J$ & $J^{P}$ & Set I & Set II & Set III & Set IV  & Ref.~\cite{Kim:2022mpa}\\
\colrule
$1^1S_0$ & $0^+$ & $3834.11$ & $3831.98$ & $3837.14$ & $3834.90$  & $4132$ \\
$1^3S_1$ & $1^+$ & $3907.65$ & $3906.38$ & $3909.15$ & $3907.59$ & $4151$ \\
$1^5S_2$ & $2^+$ & $4054.74$ & $4055.18$ & $4053.15$ & $4052.98$ & $4185$ \\
$1^1P_1$ & $1^-$ & $4335.02$ & $4342.18$ & $4341.32$ & $4347.61$ & $4423$ \\
$1^3P_0$ & $0^-$ & $4350.83$ & $4348.80$  & $4349.46$ & $4346.39$ & $4430$ \\
$1^3P_1$ & $1^-$ & $4416.72$ & $4414.12$ & $4417.13$ & $4413.93$ & $4430$ \\
$1^3P_2$ & $2^-$ & $4359.25$ & $4367.61$ & $4367.18$ & $4375.05$ & $4430$ \\
$1^5P_1$ & $1^-$ & $4379.91$ & $4379.52$ & $4380.69$ & $4379.33$ & $4442$ \\
$1^5P_2$ & $2^-$ & $4432.25$ & $4434.85$ & $4437.39$ & $4439.34$ & $4442$ \\
$1^5P_3$ & $3^-$ & $4399.57$ & $4409.28$ & $4409.58$ & $4418.93$ & $4442$ \\
$2^1S_0$ & $0^+$ & $4526.52$ & $4528.09$ & $4539.88$ & $4541.39$ & $4546$ \\
$2^3S_1$ & $1^+$ & $4558.48$ & $4559.71$ & $4570.33$ & $4571.46$ & $4560$ \\
$2^5S_2$ & $2^+$ & $4622.39$ & $4622.95$ & $4631.24$ & $4631.61$ & $4585$ \\
$2^1P_1$ & $1^-$ & $4836.59$ & $4839.76$ & $4848.30$  & $4850.10$  &        \\
$2^3P_0$ & $0^-$ & $4835.60$  & $4831.55$ & $4841.65$ & $4836.95$ &        \\
$2^3P_1$ & $1^-$ & $4889.44$ & $4885.39$ & $4897.07$ & $4892.82$ &        \\
$2^3P_2$ & $2^-$ & $4859.44$ & $4864.39$ & $4872.76$ & $4877.33$ &        \\
$2^5P_1$ & $1^-$ & $4863.13$ & $4860.77$ & $4870.73$ & $4868.33$ &        \\
$2^5P_2$ & $2^-$ & $4911.47$ & $4911.38$ & $4922.15$ & $4921.61$ &        \\
$2^5P_3$ & $3^-$ & $4896.31$ & $4902.31$ & $4910.84$ & $4916.35$ &
\end{tabular}
\end{ruledtabular}
\end{table*}

As shown in Tables~\ref{tab:m10l} and \ref{tab:m11l}, when $T_{cc}(3875)^+$ is assumed as the $J^P=1^+$ ground doubly charmed tetraquark candidate with a light ``good'' antidiquark, there exists another $J^P=0^+$ ground doubly charmed tetraquark with a light ``bad'' antidiquark. This ground doubly charmed tetraquark with isospin-$1$ (denoted with ``$T_{cc}(3834)$'') has mass $3834$ MeV in the $AL1$ ($r_c=0$) potential and $3837$ MeV in the $AL2$ ($r_c\neq 0$) potential, where the $r_c=0$ and $r_c\neq 0$ brings very small difference to the predicted mass. Though the light ``good'' antidiquark has a mass $\sim 170$ MeV lower than the light ``bad'' antidiquark, the ground $J^P=1^+$ doubly charmed tetraquark $T_{cc}(3875)^+$ with a light ``good'' antidiquark has a mass $\sim 40$ MeV higher than the $J^P=0^+$ ground doubly charmed tetraquark ``$T_{cc}(3834)$'' with a light ``bad'' antidiquark. The existence of such an ``exotic" ground tetraquark is an explicit indication of the diquark-antidiquark structure and dynamics in the doubly charmed tetraquarks.

To see the pattern of the mass spectra of doubly charmed tetraquarks, the mass splittings between different excitations are also computed and listed in Tables~\ref{tab:ms10l} and \ref{tab:ms11l}, respectively. As a comparison, the mass splittings in the chiral diquark model and the mass splittings between $D$ mesons~\cite{ParticleDataGroup:2024cfk} corresponding to the same excitations are presented in the next two columns. The mass of each multiplet is taken as the spin-weighted average mass
\begin{equation}
\overline{M_{nL}}=\frac{\sum_J{(2J+1)M_{nL_{J}}}}{\sum_J{(2J+1)}}.
\end{equation}

\begin{table*}[htb]
\caption{Mass splittings (in MeV) of radially and orbitally excited $[cc][\bar u\bar d]$ with light ``good'' antidiquark, where the parameters are chosen as Set I-IV.}
\label{tab:ms10l}
\begin{ruledtabular}
\begin{tabular}{ccccccccccc}
 & Set I & Set II & Set III & Set IV  & Ref.~\cite{Kim:2022mpa} & $D$ meson~\cite{ParticleDataGroup:2024cfk}\\
\colrule
$1P-1S$ & $419.67$ & $423.77$ & $426.15$ & $430.20$ & $292$ &$456.00$ \\
$2S-1S$ & $636.55$ & $636.39$ & $646.77$ & $646.93$ & $402$ & $634.27$ \\
$2P-1P$ & $517.56$ & $513.75$ & $523.10$ & $519.45$ &    -      & -     \\
$2P-2S$ & $300.69$ & $301.14$ & $302.48$ & $302.72$ &    -      & -
\end{tabular}
\end{ruledtabular}
\end{table*}

\begin{table*}[htb]
\caption{Mass splittings (in MeV) of radially and orbitally $[cc][\bar u\bar d]$ with light ``bad'' antidiquark, where the parameters are chosen as Set I-IV.}
\label{tab:ms11l}
\begin{ruledtabular}
\begin{tabular}{ccccccccccc}
 & Set I & Set II & Set III & Set IV  & Ref.~\cite{Kim:2022mpa} & $D$ meson~\cite{ParticleDataGroup:2024cfk}\\
\colrule
$1P-1S$ & $407.70$ & $413.05$ & $413.55$ & $418.73$ & $292$ & $456.00$ \\
$2S-1S$ & $609.24$ & $610.55$ & $619.64$ & $621.25$ & $402$ & $634.27$ \\
$2P-1P$ & $490.06$ & $487.07$ & $495.69$ & $492.83$ & -     & -        \\
$2P-2S$ & $288.52$ & $289.57$ & $289.60$ & $290.31$ & -     & -
\end{tabular}
\end{ruledtabular}
\end{table*}

For the mass splittings, the data of established $D$ mesons in PDG~\cite{ParticleDataGroup:2024cfk} is listed in Table~\ref{tab:mme}.
\begin{table}[htb]
\caption{\label{tab:mcha}
Mass spectrum (in MeV) of charmed meson from PDG~\cite{ParticleDataGroup:2024cfk}.}
\label{tab:mme}
\begin{ruledtabular}
\begin{tabular}{ccccccccccc}
State & $n^{2S+1}L_J$ & $J^{P}$ & Mass~\cite{ParticleDataGroup:2024cfk} \\
\colrule
$D^{\pm}$     & $1^1S_0$      & $0^{-}$  & $1869.66\pm0.05$ \\
$D^{0}$       & $1^1S_0$      & $0^{-}$  & $1864.84\pm0.05$ \\
$D^*(2007)^0$ & $1^3S_1$      & $1^{-}$  & $2006.85\pm0.05$ \\
$D^*(2010)^{\pm}$ & $1^3S_1$  & $1^{-}$  & $2010.26\pm0.05$ \\
$D_1(2430)^0$ & $1^1P_1$      & $1^{+}$  & $2412\pm9$   \\
$D_0^*(2300)$  & $1^3P_0$      & $0^{+}$  & $2343\pm10$  \\
$D_1((2420)^0$ & $3^1P_1$     & $1^{+}$  & $2422.1\pm0.6$  \\
$D_2^*(2460)$ & $1^3P_2$      & $2^{+}$  & $2461.1\pm0.8$  \\
$D_0(2550)^0$ & $2^1S_0$      & $0^{-}$  & $2549\pm19$   \\
$D_1^*(2600)^0$  & $2^3S_1$   & $1^{-}$  &  $2627\pm10$
\end{tabular}
\end{ruledtabular}
\end{table}

As shown in Tables~\ref{tab:ms10l} and \ref{tab:ms11l}, the $1S-1P$ and $1S-2S$ mass splittings of the doubly charmed tetraquarks are about $410-430$ MeV and $610-650$ MeV, respectively. The corresponding mass splittings of the doubly charmed tetraquarks with light ``good'' antidiquark is about $10-25$ MeV higher than those of the doubly charmed tetraquarks with light ``bad'' antidiquark. These mass splittings are similar to those for normal $D$ mesons. The $2S-2P$ mass splitting are about $120$ MeV smaller than the $1S-1P$ mass splittings, the $1P-2P$ mass splitting are also about $120$ MeV smaller than the $1S-2S$ mass splittings. In our calculation, many-body interactions and residual interaction~\cite{Lin:2024olg} have not been taken into account.

\section{summary}\label{summary}
In this work, a doubly charmed tetraquark is assumed consisting of a charmed diquark and a light antidiquark, the interactions in tetraquark are modulated as two kinds of interactions: the first type is the interaction between a quark and another quark or the interaction between an antiquark and another antiquark, and the second type is the interaction between the charmed diquark and the light antidiquark. In practical calculations, the modified $AL$-type quark-quark and quark-antiquark potentials proposed by Semay et al are employed for the interactions. Masses of some constituent diquarks and doubly charmed tetraquarks are calculated in terms of those potentials.

In the calculation of the vector charmed diquark and light antidiquark, the same parameters fitted from ordinary mesons and baryons are employed. The mass of the charmed diquark with spin-$1$ is calculated with $3499.5$ MeV and $3518.6$ MeV in the $AL1$ and $AL2$ potential, respectively. The calculated mass of the vector charmed diquark is almost the same as the predicted $3.51$ GeV in QCD sum rule.

For the light antidiquarks, there are ``good'' and ``bad'' antidiquarks. The light ``good'' antidiquark is predicted to have mass $666.2$ MeV and $673.9$ MeV in the $AL1$ and $AL2$ potential, respectively. The light ``bad'' antidiquark is predicted to have mass $834.1$ MeV and $841.8$ MeV in the $AL1$ and $AL2$ potential, respectively. The light ``good'' antidiquark has a mass $\sim 170$ MeV lower than the light ``bad'' antidiquark, which is consistent with previous phenomenological analysis: $\sim 200$ MeV. Our predicted diquark mass difference is spectator quark independent. The obtained masses of the light antidiquarks change small in different $AL$ potentials and they are much larger than the popular ones obtained in other models.

After detailed construction of the wave functions of the doubly charmed tetraquarks, we have computed their mass spectra from $S$-wave to $P$-wave excitations in terms of the fitted parameters from $T_{cc}(3875)^+$ and $X(3872)$. For the doubly charmed tetraquarks with light ``good'' antidiquark, there are one $S-$wave excitation with $J^P=1^+$, and three $P-$wave excitations with $J^P=0^-$, $J^P=1^-$ and $J^P=2^-$, respectively. For the doubly charmed tetraquarks with light ``bad'' antidiquark, there are three $S-$wave excitations with $J^P=0^+$, $J^P=1^+$ and $J^P=2^+$, respectively. Further more, there are seven $P-$wave excitations with $J^P=0^-$ (one), $J^P=1^-$ (three), $J^P=2^-$ (two) and $J^P=3^-$ (one), respectively. These features of the emergence of tetraquarks are explicit indication of the quark structure in tetraquarks.

When $T_{cc}(3875)^+$ is assumed as the $J^P=1^+$ ground doubly charmed tetraquark candidate with isospin-$0$ made from a light ``good'' antidiquark. A $IJ^P=10^+$ ground doubly charmed tetraquark with a light ``bad'' antidiquark inside is also expected. This $IJ^P=10^+$ ground tetraquark has mass about $3831-3837$ MeV in the $AL$ potentials, which exhibits an explicit indication of the diquark-antidiquark structure and dynamics in the doubly charmed tetraquarks. The next $P-$wave excited doubly charmed tetraquarks corresponding to $T_{cc}(3875)^+$ are predicted to have masses with an average $410-430$ MeV higher. These numerical results can serve as a guide for experiments in the search of tetraquarks, and we expect the $IJ^P=10^+$ ground doubly charmed tetraquark ``$T_{cc}(3834)$'' to be found in forthcoming experiments.

The mass splittings among doubly charmed tetraquarks in different excitations are similar to those for ordinary $D$ mesons. The $1S-1P$ and $1S-2S$ mass splittings of the doubly charmed tetraquarks are about $410-430$ MeV and $610-650$ MeV, respectively. The corresponding mass splittings of the doubly charmed tetraquarks with light ``good'' antidiquark is about $10-25$ MeV higher than those of the doubly charmed tetraquarks with light ``bad'' antidiquark from $1S$ to $2P$ excitations. The $2S-2P$ mass splitting are about $120$ MeV smaller than the $1S-1P$ mass splittings, the $1P-2P$ mass splitting are also about $120$ MeV smaller than the $1S-2S$ mass splittings. The pattern of mass splittings in tetraquarks exhibits the properties of quark interactions, and could serve as a support of the constituent diquark-antidiquark potential model.

The effects of mixing, many-body interactions and cluster structure etc in doubly charmed tetraquarks have not been taken into account explicitly, the isospin-dependent interaction has not been taken into account either. The many-body interactions and cluster effects are possible small if the quarks and antiquarks in tetraquarks make point-like compact cluster in real world. The parameters in the $AL$ potentials fitted by $T_{cc}(3875)^+$ and $X(3872)$ may bring in an uncertainty to the predicted mass spectra of doubly charmed tetraquarks if these two states are not the assumed ones. However, through a simple uncertainties estimate, those uncertainties of mass spectra are not large. Of course, if more doubly or hidden charmed tetraquarks are observed and definitely identified, the experimental data can be employed for the fitting procedure to improve our predictions.
\begin{acknowledgments}
This work is supported by National Natural Science Foundation of China under the grant No. 11975146.
\end{acknowledgments}


\end{document}